\documentclass[11pt,preprint]{aastex}
\usepackage{epsfig}

\newcommand       \Angstrom     {\,{\rm \AA}} 
\newcommand       \AU           {\,{\rm AU}}          

\newcommand       \cm           {\,{\rm cm}}

\newcommand	  \g		{\,{\rm g}}
\newcommand       \K            {\,{\rm K}}
\newcommand	  \pc		{\,{\rm pc}}

\newcommand	  \myr		{\,{\rm Myr}}
\newcommand       \simlt        {\lesssim}

\newcommand       \gtsim        {\gtrsim}

\newcommand       \mum          {\,{\rm \mu m}}
\newcommand	  \ppm		{\,{\rm ppm}}
\newcommand	  \Teff	        {T_{\rm eff}}
\newcommand	  \bp	        {\beta\ {\rm Pictoris}}
\newcommand	  \hra	        {{\rm HR\,4796A}}
\newcommand	  \hrb	        {{\rm HR\,4796B}}
\newcommand	  \amin	        {a_{\rm min}}
\newcommand	  \amax	        {a_{\rm max}}
\newcommand	  \rin	        {r_{\rm in}}
\newcommand	  \rout	        {r_{\rm out}}
\newcommand	  \xsun         {\left[{\rm X/H}\right]_{\odot}}
\newcommand	  \csun         {\left[{\rm C/H}\right]_{\odot}}
\newcommand	  \nsun         {\left[{\rm N/H}\right]_{\odot}}
\newcommand	  \osun         {\left[{\rm O/H}\right]_{\odot}}
\newcommand	  \fesun        {\left[{\rm Fe/H}\right]_{\odot}}
\newcommand	  \mgsun        {\left[{\rm Mg/H}\right]_{\odot}}
\newcommand	  \sisun        {\left[{\rm Si/H}\right]_{\odot}}
\newcommand	  \xdust        {\left[{\rm X/H}\right]_{\rm dust}}
\newcommand	  \cdust        {\left[{\rm C/H}\right]_{\rm dust}}
\newcommand	  \ndust        {\left[{\rm N/H}\right]_{\rm dust}}
\newcommand	  \odust        {\left[{\rm O/H}\right]_{\rm dust}}
\newcommand	  \fedust       {\left[{\rm Fe/H}\right]_{\rm dust}}
\newcommand	  \mgdust       {\left[{\rm Mg/H}\right]_{\rm dust}}
\newcommand	  \sidust       {\left[{\rm Si/H}\right]_{\rm dust}}
\newcommand	  \osil         {\left[{\rm O/H}\right]_{\rm sil}}
\newcommand	  \owater       {\left[{\rm O/H}\right]_{\rm water}}
\newcommand	  \xgas         {\left[{\rm X/H}\right]_{\rm gas}}
\newcommand	  \cgas         {\left[{\rm C/H}\right]_{\rm gas}}
\newcommand	  \ngas         {\left[{\rm N/H}\right]_{\rm gas}}
\newcommand	  \ogas         {\left[{\rm O/H}\right]_{\rm gas}}

\newcommand	  \mux          {\mu_{\rm X}}
\newcommand	  \muc          {\mu_{\rm C}}
\newcommand	  \muh          {\mu_{\rm H}}
\newcommand	  \mun          {\mu_{\rm N}}
\newcommand	  \muo          {\mu_{\rm O}}
\newcommand	  \mufe         {\mu_{\rm Fe}}
\newcommand	  \mumg         {\mu_{\rm Mg}}
\newcommand	  \musi         {\mu_{\rm Si}}
\newcommand	  \msil         {m_{\rm sil}} 
\newcommand	  \mcarb        {m_{\rm carb}} 
\newcommand	  \mice         {m_{\rm ice}} 
\newcommand	  \mdust        {m_{\rm dust}} 
\newcommand{\figwidth}{3.5in}

\countdef\decade=200
\advance\decade by \year
\countdef\hours=201
\advance\hours by \time
\divide\hours by 60
\countdef\mins=202
\advance\mins by \hours
\multiply\mins by 60
\multiply\hours by 100
\countdef\miltime=203
\advance\miltime by \hours
\advance\miltime by \time
\advance\miltime by -\mins

\shorttitle{The Infrared Spectral Energy Distribution}

\begin{document}

\title{
 \vspace*{-2.0em}
  {\normalsize\rm submitted to {\it The Astrophysical Journal Letters}}\\
  \vspace*{1.0em}
What Can We Learn from the Infrared Spectral Energy Distributions of 
Dust Disks? 
	 }

\author{Aigen Li and J.I. Lunine}
\affil{Theoretical Astrophysics Program,
       University of Arizona, Tucson, AZ 85721;\\
        {\sf agli@lpl.arizona.edu, jlunine@lpl.arizona.edu}}

\begin{abstract}
The spectral energy distributions (SEDs) of dust disks are widely
used to infer dust properties (compositions and sizes) and
disk structures (dust spatial distributions) which might be 
indicative of the presence or absence of planets or smaller
bodies like asteroids and comets in the disk. Based on 
modelling of the SED of $\hra$, a young main-sequence
star with the largest fractional infrared (IR) emission, we show
that the SED {\it alone} is not a good discriminator of dust 
size, spatial distribution (and composition if no spectroscopic
data are available). A combination of SED, mid-IR spectroscopy,
and coronagraphic near-IR imaging of scattered starlight and 
mid-IR imaging of dust thermal emission provides a better 
understanding of these properties.
\end{abstract}


\section{Introduction\label{sec:intro}}
Over the past 2 decades, impressive evidence has been assembled 
for the existence of circumstellar dust disks around main-sequence (MS)
stars as well as pre-MS stars (T Tauri stars and Herbig Ae/Be stars),
post-MS stars (red giants), and a white dwarf (see Zuckerman 2001 for 
a review). A wide variety of observational techniques have been
employed to study the formation/evolution and physical/chemical
properties of dust disks: optical and near-infrared (IR) imaging of
scattered stellar light, photometric measurements of dust thermal emission
from near-IR to submillimeter, spectroscopic observations of mid-IR 
dust emission features and gas emission lines and ultraviolet (UV) 
and visible gas absorption lines. The spectral energy distribution   
(SED) is of particular interest in inferring the size and composition of
dust grains and the disk structure (dust spatial distribution). 
However, the limitations of this method have not been adequately explored;
for example, in modelling the $\bp$ SED, Li \& Greenberg (1998) found 
that the dust spatial distribution is coupled with the distribution of
dust sizes, i.e., the distribution of dust in the disk and dust sizes 
cannot be uniquely determined simultaneously by the SED alone. 

It is the aim of this {\it Letter} to quantify the limitations on 
the information derived from the SED modelling concerning the dust
properties (composition and size) and disk structure. 
We take the SED of $\hra$, a nearby (distance to Earth $d\approx 67\pm 3\pc$) 
young MS star (age $\approx 8\pm 3\myr$) of spectral type A0\,V
(effective temperature $\Teff\approx 9500\K$) for comparison with 
model results. $\hra$ has the largest fractional IR luminosity 
relative to the total stellar luminosity
($L_{\rm IR}/L_{\star}\approx 5\times 10^{-3}$)
among the $\sim 1500$ A-type MS stars in the {\it Yale Bright Star
Catalogue} (Jura 1991). The dust disk in orbit around $\hra$ has
been extensively studied, both observationally and theoretically
(see Zuckerman 2001 and references therein). With its relatively
well determined dust and disk properties, $\hra$ serves as a good
comparison basis for the SED modelling efforts described in this 
{\it Letter}. We stress that the main purpose of this {\it Letter} 
is not to carry out a detailed study of the $\hra$ dust disk 
which we defer to a subsequent paper (A. Li \& J.I. Lunine 2002,
in preparation). We will first outline our approach in \S\ref{sec:model}.
We then discuss in \S\ref{sec:spasize} the degeneracy between 
the dust spatial distribution and dust sizes under the assumption of
pure silicate dust. In \S\ref{sec:composition} we show that the dust 
composition is not well constrained by the observed SED unless mid-IR 
spectroscopic dust emission features are available. 
We discuss in \S\ref{sec:discussion} the possible dust composition 
and morphology from the evolutionary point of view that circumstellar 
dust disks around (pre-)MS stars are formed through the coagulation 
of interstellar solids. We also summarize our major conclusions 
in \S\ref{sec:discussion}.

\section{Modelling the Dust IR Emission\label{sec:model}}
Grains in the optically thin dust disk of $\hra$ absorb the stellar 
UV/optical photons and then re-radiate them in the IR. Very small grains 
or large molecules (with spherical radius smaller than 100$\Angstrom$)
which are subject to single-photon heating (Draine \& Li 2001) will not be 
considered here although their presence in the disk can not be ruled 
out (e.g., polycyclic aromatic hydrocarbon molecules are seen in the
disk of HD\,100546 through their 3.3, 6.2, 7.7, 8.6, and 11.3$\mum$
emission features [Malfait et al.\ 1998]). 
For grains $\gtsim 100\Angstrom$ we first use Mie theory to calculate 
absorption cross sections assuming a spherical shape. We then 
calculate grain equilibrium temperatures (and corresponding 
IR emission) by balancing absorption and emission.
For a given dust size distribution and a given disk structure (dust 
spatial density distribution), the emergent IR emission spectrum can 
be obtained by integrating over the dust size range, 
and over the entire disk. 

We have the following parameters to be specified or constrained:
(1) dust composition -- we will consider amorphous silicate, amorphous
carbon, and cometary grains (porous aggregates of small silicate and
carbon dust; Greenberg 1982); (2) dust sizes -- we will
consider a power law dust size distribution $dn(a)/da \propto a^{-\alpha}$ 
characterized by the lower-cutoff $\amin$, upper-cutoff $\amax$ 
and power-law index $\alpha$ (where $a$ is the spherical radius);
(3) dust spatial density distribution -- we will consider two 
dramatically different distribution functions:
a power law distribution $dn(r)/dr \propto r^{-\beta}$ 
and a Gaussian distribution 
$dn(r)/dr \propto \exp[-4\ln2\{(r-r_0)/\Delta\}^2]$, 
the former is characterized by the disk's inner boundary $\rin$, 
outer boundary $\rout$, and power-law index $\beta$;
the latter is characterized by the radial position $r_0$ 
where $dn(r)/dr$ peaks and the full width half maximum (FWHM) $\Delta$.

Jura et al.\ (1995, 1998) and Augereau et al.\ (1999) estimated 
that grains with radius exceeding a few micron are stable against
radiation pressure. Telesco et al.\ (2000) estimated the 
``characteristic'' radius for the 10--20$\mum$ mid-IR-emitting 
grains to be $\approx 1-1.5\mum$. We therefore adopt $\amin=1\mum$. 
We take $\amax=1\cm$ (this is a noncritical parameter since grains 
larger than $\sim 100\mum$ are like blackbodies
and their IR emission spectra are size-insensitive).

To be conservative, we first set the inner boundary at
$\rin=0.15\AU$ inside of which micron-sized silicate and 
carbonaceous grains would be heated to $\gtsim 1500\K$ and evaporate.
The outer boundary is taken to be $\rout=250\AU$ which is 
expected from the disk truncation caused by the tidal effects 
of $\hrb$, a companion star of $\hra$ (Jayawardhana et al.\ 1998).    
Other values for $\rin$ and $\rout$ estimated from 
the near- and mid-IR imaging observations will also be discussed
(see \S\ref{sec:spasize}). For the Gaussian-type dust spatial
distribution, we take the peak distance (from the central star)
of the distribution to be $r_0=70\AU$ as reflected from the near-IR 
imaging of scattered starlight (Schneider et al.\ 1999) and mid-IR 
imaging of dust thermal emission (Jayawardhana et al.\ 1998; 
Koerner et al.\ 1998; Telesco et al.\ 2000). Therefore, we are only 
left with two free parameters: the dust size distribution power index
$\alpha$ and the dust spatial distribution power index $\beta$ or
the FWHM $\Delta$ of the Gaussian-type spatial distribution.  
   
\section{Dust Spatial Distributions and Dust Sizes\label{sec:spasize}}
We first consider compact silicate grains. We adopt the indices of
refraction of the Draine \& Lee (1984) ``astronomical'' silicate.
We approximate the $\hra$ radiation field by the Kurucz model 
atmosphere spectrum for A0\,V stars (Kurucz 1979). We will compare
our model results with the available photometric data compiled by
Augereau et al.\ (1999) for the $\hra$ disk. 

\begin{figure}[h]
\begin{center}
\epsfig{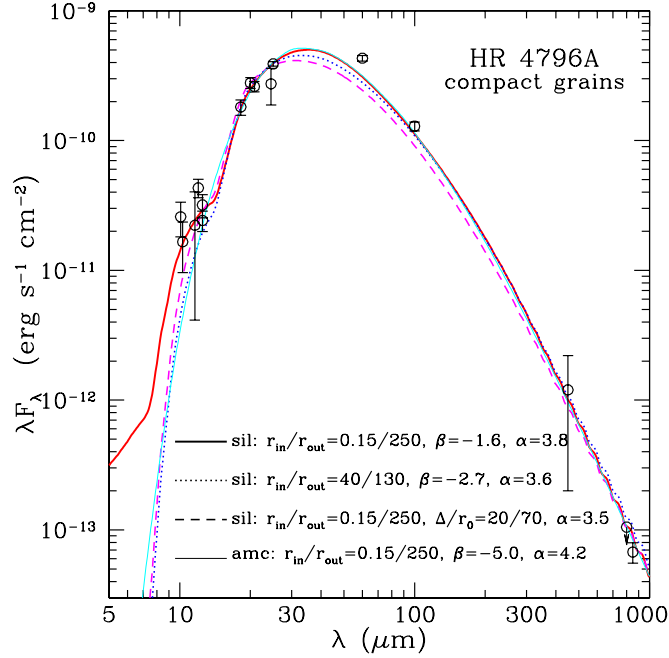}
\end{center}\vspace*{-1em}
\caption{
        \label{fig:silamc}
        \footnotesize
        Comparison of the observed spectral energy distribution
        of the $\hra$ dust disk to theoretical IR emission spectra
        calculated from (1) heavy solid line -- pure compact silicate grains 
        (mass density $\rho_{\rm sil}\approx 3.5\g\cm^{-3}$)
        with a power law size distribution $dn(a)/da \propto a^{-\alpha}$
        where $\alpha=3.8$ and $a\in [1\mum,1\cm]$, 
        and a power-law spatial distribution 
        $dn(r)/dr \propto r^{-\beta}$ where $\beta=-1.6$ 
        ($r\in [0.15\AU,250\AU]$) and total dust mass 
        $\mdust\approx 4.98\times10^{26}\g$;
        the slope change at $\lambda \sim 7\mum$ is due to the
        onset of the 10$\mum$ Si-O silicate vibrational mode;
        the little bump at 10$\mum$ is the Si-O silicate band;  
        (2) dotted line -- silicate grains with a flatter power-law size 
        distribution of $\alpha=3.6$ ($a\in [1\mum,1\cm]$) 
        and a steeper and less extended spatial distribution 
        of $\beta=-2.7$ ($r\in [40\AU,130\AU]$)
        and $\mdust\approx 4.53\times10^{26}\g$;
        (3) dashed line -- silicate grains with a power-law size 
        distribution of $\alpha=3.5$ ($a\in [1\mum,1\cm]$) 
        and a Gaussian spatial distribution of
        $dn(r)/dr \propto \exp[-4\ln2\{(r-r_0)/\Delta\}^2]$ 
        with its peak at $r_0=70\AU$ and a FWHM $\Delta=20\AU$
        ($r\in [0.15\AU,250\AU]$)
        and $\mdust\approx 3.46\times10^{26}\g$;
        (4) thin solid line -- pure compact amorphous carbon grains 
        (mass density $\rho_{\rm amc}\approx 1.8\g\cm^{-3}$) with
        a power law size distribution of $\alpha=4.2$ ($a\in [1\mum,1\cm]$) 
        and a power-law spatial distribution of 
        $\beta=-5.0$ ($r\in [0.15\AU,250\AU]$)
        and $\mdust\approx 1.89\times10^{26}\g$.
        }
\end{figure}

Assuming a power-law dust spatial distribution for
$\rin=0.15\AU$ and $\rout=250\AU$, the best fit to the
observed SED is provided by $\alpha\approx 3.8$ and $\beta\approx -1.6$
(see Figure \ref{fig:silamc}) -- the density of dust {\it increases} 
outward from the star through the whole disk. Since various studies
suggest that the $\hra$ disk has an inner hole at $r\sim 40-60\AU$
and an outer edge sharply truncated at $\sim 80-130\AU$
(Jura et al.\ 1993; Jura et al.\ 1995; Jayawardhana et al.\ 1998; 
Koerner et al.\ 1998; Schneider et al.\ 1999; Wyatt et al.\ 1999;
Telesco et al.\ 2000), we also consider models with $\rin=40\AU$ 
and $\rout=130\AU$. To obtain a satisfactory fit to the observed SED, 
an even steeper outward increase of dust distribution is needed 
(for enhancing the amount of warm dust). 
In Figure \ref{fig:silamc} we also present 
the best-fit model spectrum for $\rin=40\AU$ and $\rout=130\AU$ 
obtained from $\alpha \approx 3.6$ and $\beta\approx -2.7$.
We have also modelled the $\hra$ SED in terms of broken power-laws
for both the dust spatial distribution and dust sizes:
$dn(r)/dr \propto r^{-\beta_1}$, $dn(a)/da \propto a^{-\alpha_1}$
for $\rin \simlt r \simlt r_0$ and
$dn(r)/dr \propto r^{-\beta_2}$, $dn(a)/da \propto a^{-\alpha_2}$
for $r_0 \simlt r \simlt \rout$. These 2-power-law models also
require $\beta < 0$ and we do not see significant improvements
compared with single-power-law models.

Motivated by the {\it NICMOS} (Near-IR Camera and Multi-Object
Spectrometer) discovery that the surface brightness of the $\hra$
disk's scattered light sharply peaks at $\sim 70\AU$ 
(Schneider et al.\ 1999), we adopt a Gaussian function peaking 
at $r_0=70\AU$ for the dust spatial distribution. 
As long as the FWHM $\Delta \simlt 30\AU$, the bulk of the dust 
lies at $40\simlt r\simlt 130\AU$ and we expect little difference
between the $[\rin=0.15\AU, \rout=250\AU]$ model and
the $[\rin=40\AU, \rout=130\AU]$ model. 
Therefore, for the Gaussian-type distribution, 
we only consider $\rin=0.15\AU$ and $\rout=250\AU$. 
Figure \ref{fig:silamc} plots the best-fit spectrum calculated
from $\rin=0.15\AU$, $\rout=250\AU$, $r_0=70\AU$, $\Delta=20\AU$,
and $\alpha=3.5$. This appears to be consistent with the {\it NICMOS}
detection of a sharply bounded and narrow ($\Delta \simlt 17\AU$)
ring-like disk (Schneider et al.\ 1999). 

It is seen in Figure \ref{fig:silamc} that, although the dust spatial
distributions are strikingly different, all three models are able to
provide similarly good fits to the observed SED except the IRAS 
[{\it Infrared Astronomical Satellite}] 60$\mum$ data (we note that 
the IRAS photometric uncertainty given by Augereau et al.\ (1999)
might have been underestimated [Beichman et al.\ 1989]), provided 
that the dust sizes are different. 

\section{Dust Composition\label{sec:composition}}
We have seen in \S\ref{sec:spasize} that pure solid silicate 
grains with various sizes and spatial distributions are successful
in reproducing the SED of $\hra$. Similarly, the $\hra$ SED can also 
be fitted by pure solid amorphous carbon grains using the index of 
refraction of Rouleau \& Martin (1991). For illustration, we present 
in Figure \ref{fig:silamc} the best-fit single-power-law model spectrum 
calculated from $\rin=0.15\AU$, $\rout=250\AU$, $\beta = -5.0$, 
and $\alpha=4.2$. Although it is unlikely that (proto)planetary disk
dust is mainly made of carbonaceous material, the observed SED {\it alone}
is unable to rule out the pure carbon dust model unless the 9.7$\mum$
Si-O and 18$\mum$ O-Si-O features are detected.\footnote{%
 The 8--13$\mum$ spectroscopic observations of the $\hra$ disk show 
 that the thermal emission in the silicate feature is very weak
 (Sitko, Lynch, \& Russell 2000).
 }
We will discuss this further in \S\ref{sec:discussion}.

\begin{figure}[h]
\begin{center}
\epsfig{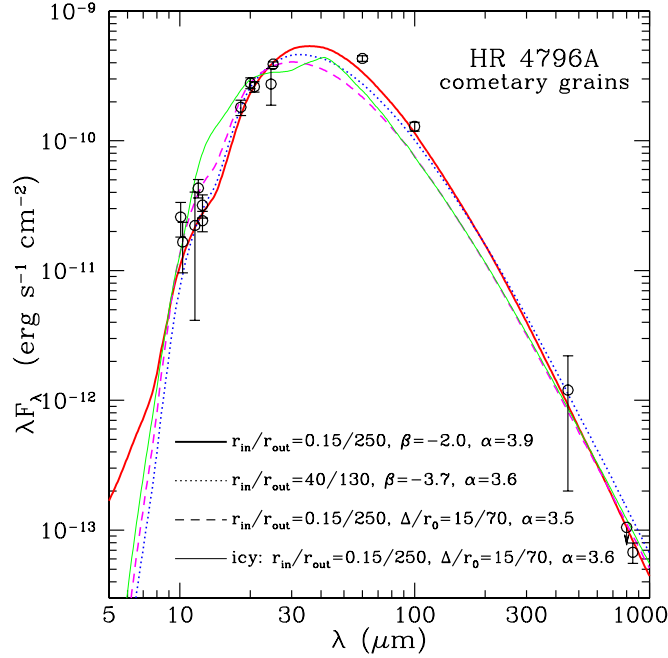}
\end{center}\vspace*{-1em}
\caption{
        \label{fig:comet}
        \footnotesize
        Comparison of the observed SED of the $\hra$ dust disk to 
        theoretical IR emission spectra of cometary grains (composed 
        of silicate, carbonaceous dust and vacuum):
        (1) heavy solid line -- for a power law size distribution of 
        $\alpha=3.9$ ($a\in [1\mum,1\cm]$) and a power-law spatial 
        distribution of $\beta=-2.0$ ($r\in [0.15\AU,250\AU]$)
        and total dust mass $\mdust\approx 3.23\times10^{26}\g$;
        (2) dotted line -- for a power law size distribution of 
        $\alpha=3.6$ ($a\in [1\mum,1\cm]$) and a power-law spatial 
        distribution of $\beta=-3.7$ ($r\in [40\AU,130\AU]$)
        and $\mdust\approx 4.47\times10^{26}\g$;
        (3) dashed line -- for a power law size distribution of 
        $\alpha=3.5$ ($a\in [1\mum,1\cm]$) and a Gaussian spatial 
        distribution of $r_0=70\AU$ and $\Delta=15\AU$ 
        ($r\in [0.15\AU,250\AU]$) and $\mdust\approx 3.00\times10^{26}\g$.
        In all models the volume fractions of the silicate, carbonaceous, 
        and vacuum components are taken to be 0.2, 02., and 0.6, 
        respectively (effective mass density $\approx 1.06\g\cm^{-3}$).
        Also plotted is the model spectrum (thin solid line) calculated 
        from a Gaussian distribution ($\rin=0.15\AU$, $\rout=250\AU$, 
        $r_0=70\AU$, $\Delta=15\AU$) of grains with compositions
        derived in the context of the formation and evolution of
        dust disks ($\mcarb/\msil = 0.7$ and a porosity of $P=0.6$ 
        for $r<30\AU$; compact icy grains with $\mcarb/\msil = 0.7$
        and $\mice/[\msil+\mcarb] = 0.8$ for $r>30\AU$;
        see \S\ref{sec:discussion}) with a power-law size distribution 
        of $\alpha=3.6$ and total dust mass 
        $\mdust\approx 1.94\times10^{26}\g$.
        }
\end{figure}

Jura et al.\ (1998) argued that the $\hra$ disk are composed of
cometary icy grains. The $\hra$ SED was closely reproduced by 
Augereau et al.\ (1999) in terms of a cold interstellar dust component
and a hot cometary dust component. 
In this Section we also model the $\hra$ SED by cometary grains. 
Following Greenberg (1982), we model cometary dust as fluffy aggregates 
of interstellar silicate and carbonaceous grains.\footnote{%
 In the framework of the Greenberg comet model (Greenberg 1982, 1998;
 Greenberg \& Li 1999), the silicate dust and carbon dust are physically
 related through a core-mantle structure. In this work the physical
 relationship between the silicate dust and carbon dust does not matter
 since the Bruggeman effective medium theory employed to calculate
 dust optical properties does not distinguish inclusions from matrix
 (Bohren \& Huffman 1983). Therefore, the Greenberg comet model is 
 not just limited to the core-mantle interstellar dust model (Li \&
 Greenberg 1997), but also applicable to other popular dust models
 such as the separate silicate/graphite model (Mathis, Rumpl, \& 
 Nordsieck 1997, Draine \& Lee 1984, Weingartner \& Draine 2001, 
 Li \& Draine 2001) and the composite dust model (Mathis 1996).
 However, in the dense protostellar environment, it is possible 
 that graphite grains may be destroyed by chemical attacks of
 O, H$_2$O (Draine 1985).         
 }   
We take the grain porosity (the volume fraction of vacuum) to be 0.6
(Augereau et al.\ 1999). The volume ratio of the silicate component to
the carbonaceous component is taken to be 1.0 as derived from the in situ 
measurement of the coma dust of comet Halley (Kissel \& Krueger 1987).
We use Mie theory and the Bruggeman effective medium theory 
(Bohren \& Huffman 1983) to calculate the absorption cross sections
for porous cometary grains. Similar to the pure solid silicate dust
model (see \S\ref{sec:spasize}), the cometary dust model is also able
to provide reasonably good fits to the $\hra$ SED with various dust
sizes and dust spatial distributions. In Figure \ref{fig:comet} we 
show the model spectra calculated from cometary grains with a power-law
size distribution and a power-law or Gaussian spatial distribution:
(1) $\alpha=3.9$, $\beta=-2.0$ and $r\in [0.15\AU,250\AU]$;
(2) $\alpha=3.6$, $\beta=-3.7$ and $r\in [40\AU,130\AU]$;
(3) $\alpha=3.5$, $r_0=70\AU$, $\Delta=15\AU$ and $r\in [0.15\AU,250\AU]$.

\section{Discussions\label{sec:discussion}}
It is shown in \S\ref{sec:spasize} and \S\ref{sec:composition}
that models with various compositions, sizes, and spatial distributions
are able to reproduce the observed SED of the $\hra$ disk reasonably well.
It was also shown by Jura et al.\ (1998) that icy grains with a typical
radius near 100$\mum$ are able to explain the $\hra$ SED. 
Dust thermal emission depends on its absorption and emission 
properties which are determined by its size and optical properties. 
It would not be surprising for a wide range of dust materials with
properly chosen sizes to be able to fit the SED. However, we should
not be too pessimistic: such spectrum will be useful when combined
with other constraints on the composition of the dust in circumstellar 
disks around (pre-)MS stars. The coagulation of interstellar grains that
results in fluffy and inhomogeneous aggregates occurs in 
cold, dense molecular clouds and protostellar and protoplanetary 
dust disks and plays an important role in the 
formation of planetary systems (Weidenschilling \& Cuzzi 1993). 
We can therefore approximately derive the proportional composition 
of the dust in circumstellar disks from the abundances of 
the condensable elements (C, N, O, Si, Fe, and Mg),\footnote{%
 Some H will be present, mostly in combination with O, C, and N.
 } 
assuming protostellar activities impose little modification on
protostellar grain compositions (see Beckwith, Henning, \& Nakagawa 2000). 

Let $\xsun$ be the cosmic abundance of X relative to H (we assume 
the cosmic elemental abundances are those of the solar values:
$\csun \approx 391$ parts per million (ppm),
$\nsun \approx 85.2\ppm$, $\osun \approx 545\ppm$,
$\mgsun \approx 34.5\ppm$, $\fesun \approx 34.4\ppm$,
and $\sisun \approx 28.1\ppm$ [Sofia \& Meyer 2001]);
$\xgas$ be the amount of X in gas phase
($\cgas\approx 140\ppm$, $\ngas\approx 61\ppm$, 
$\ogas\approx 310\ppm$; Fe, Mg and Si are highly depleted in dust;
see Li \& Greenberg 1997 and references therein);
$\xdust$ be the amount of X relative H locked up in dust
($\cdust = \csun-\cgas \approx 251\ppm$, $\ndust\approx 24.2\ppm$,
$\odust\approx 235\ppm$, $\mgdust\approx 34.5\ppm$,
$\fedust\approx 34.4\ppm$, $\sidust\approx 28.1\ppm$).
Assuming a stoichiometric composition of MgFeSiO$_4$ for 
interstellar silicates, the total mass of silicate dust per H atom is
$\msil \approx \fedust\mufe + \mgdust\mumg + \sidust\musi + \osil\muo
\approx 5.61\times 10^{-3}\,\muh$ 
where $\mux$ is the atomic weight of X in unit of 
$\muh\approx 1.66\times 10^{-24}\g$,
and $\osil\approx 4\,(\fedust + \mgdust + \sidust)/3\approx 129\ppm$  
is the amount of O in silicate dust per H atom (i.e., we assign 
4 O atoms for the average of the Fe, Mg, and Si abundances).
The carbonaceous dust component is dominated by C, 
with little H, N, and O (we assume H/C=0.5, O/C=0.1). 
The total mass of carbon dust per H atom is 
$\mcarb \approx \cdust\muc + \ndust\mun + 0.5 \cdust\muh + 0.1 \cdust\muo
\approx 3.88\times 10^{-3}\,\muh$. 
The C, O, and N atoms left over after accounting for the silicate
and carbon dust components are assumed to condense in icy grains 
in the form of H$_2$O, NH$_3$, CO, CO$_2$, CH$_3$OH and CH$_4$ 
(following Greenberg [1998], we assume 
CO:CO$_2$:CH$_3$OH:CH$_4$:H$_2$CO=10:4:3:1:1).
The total mass of icy grains per H atom is 
$\mice \approx \mice^{\rm C} + \mice^{\rm N} + \mice^{\rm water}$,
where the mass of C-containing ice $\mice^{\rm C}\approx \cgas\muc 
+ \cgas\,(22\muo+18\muh)/19\approx 2.87\times10^{-3}\,\muh$;
the mass of NH$_3$ ice $\mice^{\rm N}\approx \ngas(\mun+3\muh)
\approx 4.10\times 10^{-4}\,\muh$; 
the mass of water ice $\mice^{\rm water}\approx \owater(\muo+2\muh)
\approx 4.12\times 10^{-3}\,\muh$;
$\owater \approx \osun-\osil-0.1\cdust-22\cgas/19 \approx 229\ppm$ 
is the amount of O locked up in H$_2$O ice (we assume H$_2$O contains
all the remaining available O).

Therefore, as a first approximation, we may assume a mixing ratio
of $\mcarb/\msil \approx 0.7$ and $\mice/(\msil+\mcarb) \approx 0.8$
for cold regions (for hot regions where ices sublimate the dust can be
simply modelled as porous aggregates of silicate and carbon particles with
$\mcarb/\msil \approx 0.7$). This does not deviate much from the in situ 
measurements of cometary dust
($\mcarb/\msil \approx 0.5$, $\mice/[\msil+\mcarb] \approx 1.0$;
see Greenberg \& Li 1999 and references therein) which is often 
suggested as porous aggregates of unaltered interstellar dust
(Greenberg 1982; Greenberg \& Li 1999).
The porosity is a free parameter ranging from that of diffuse cloud 
interstellar dust ($\sim 0.45$, Mathis 1996) to that of very fluffy 
cometary dust ($\gtsim 0.9$, Greenberg \& Li 1999).
For illustration, we plot in Figure \ref{fig:comet} the model
spectrum calculated from a Gaussian distribution 
($\rin=0.15\AU$, $\rout=250\AU$, $r_0=70\AU$, $\Delta=15\AU$)
of grains with (1) a power law size distribution 
($\amin=1\mum$, $\amax=1\cm$, $\alpha=3.6$),
(2) $\mcarb/\msil = 0.7$ and a porosity of $P=0.6$ for $r<30\AU$;
and (3) compact icy grains with $\mcarb/\msil = 0.7$
and $\mice/(\msil+\mcarb) = 0.8$ (porous grains of $P=0.6$ become
compact after filled with ices of an amount of 
$\mice/[\msil+\mcarb] = 0.8$) for $r>30\AU$.
This will be discussed in detail in a subsequent paper
(A. Li \& J.I. Lunine 2002, in preparation).

The fact that pure amorphous carbon grains are also able to 
account for the observed SED (see \S\ref{sec:composition})
reinforces the importance of combining observations with 
theoretical calculations of dust composition in the context 
of the formation and evolution of dust disks. 
We note that the non-detection of the silicate emission features
in the $\hra$ disk (Sitko et al.\ 2000) does not necessarily 
imply the predominance of non-silicate dust in the disk.
It may just imply the lack of small and hot silicate grains.
On the other hand, the contradistinction between the various 
dust spatial distributions, which all provide a reasonably good 
fit to the observed SED (see \S\ref{sec:spasize} and 
\S\ref{sec:composition}), indicates the importance of direct
disk imaging.

In summary, the spectral energy distributions of dust disks 
{\it alone} are not necessarily able to constrain the dust compositions, 
sizes, and spatial distributions. The dust spatial and size 
distributions are coupled. Caution should be taken in discussing
the presence/absence of planets, comets, and asteroids
in the disk {\it solely} based on the observed SED. 
We argue that grains in circumstellar disk around (pre-)MS stars
are composed of silicate, carbonaceous dust (and ices in cold 
regions) and vacuum with a mixing ratio of $\mcarb/\msil \approx 0.7$ 
and $\mice/(\msil+\mcarb) \approx 0.8$. 
A combination of compositional considerations, SED, 
mid-IR spectroscopy, coronagraphic near-IR imaging of 
scattered starlight and mid-IR imaging of dust thermal emission 
will allow us to better understand the properties of circumstellar
dust disks.

\acknowledgments
We dedicate this paper to Prof. J. Mayo Greenberg 
who passed away on November 29, 2001.
A. Li thanks the University of Arizona for the ``Arizona 
Prize Postdoctoral Fellowship in Theoretical Astrophysics''.
This research was supported in part by NASA grant NAG5-10450.

\end{document}